%% file: main.tex
\documentclass[runningheads]{llncs}
\usepackage{graphicx}
\usepackage{algorithm}
\usepackage{amsmath}
\usepackage{subfigure}
\usepackage[noend]{algpseudocode}
\usepackage{float}
\usepackage{stackengine}
\usepackage{multirow}
\usepackage{booktabs}
\usepackage{tikz}
\usepackage{amssymb}
\usepackage{tabularx}
\usepackage{arydshln}

\usepackage{xcolor}

\usepackage[hidelinks,colorlinks=true,linkcolor=blue,citecolor=blue]{hyperref}
\usepackage{makecell} 

\usepackage[capitalize]{cleveref}
\crefname{section}{Sec.}{Secs.}
\Crefname{section}{Section}{Sections}
\Crefname{table}{Table}{Tables}
\crefname{table}{Tab.}{Tabs.}

\begin{document}

\title{FD-SOS: Vision-Language Open-Set Detectors for Bone Fenestration and Dehiscence Detection from Intraoral Images}

\newcommand{\repeatthanks}{\textsuperscript{\thefootnote}}

\titlerunning{FD-SOS}

\author{Marawan Elbatel\inst{1}
\and
Keyuan Liu\inst{2} \and 
Yanqi Yang\inst{2}\thanks{Corresponding authors: \email{eexmli@ust.hk}, \email{yangyanq@hku.hk}} \and Xiaomeng Li\inst{1,3}\repeatthanks
 }

\authorrunning{M. Elbatel et al.} 

\institute{Department of Electronic and Computer Engineering, The Hong Kong University of Science and Technology, Hong Kong SAR, China  \\
  \and Division of Paediatric Dentistry and Orthodontics, Faculty of Dentistry, The University of Hong Kong, Hong Kong SAR, China \\
  \and HKUST Shenzhen-Hong Kong Collaborative Innovation Research Institute, Shenzhen, China
}

\maketitle

\begin{abstract}

Accurate detection of bone fenestration and dehiscence (FD) is crucial for effective treatment planning in dentistry. While cone-beam computed tomography (CBCT) is the gold standard for evaluating FD, it comes with limitations such as radiation exposure, limited accessibility, and higher cost compared to intraoral images. In intraoral images, dentists face challenges in the differential diagnosis of FD. This paper presents a novel and clinically significant application of FD detection solely from intraoral images. To achieve this, we propose FD-SOS, a novel open-set object detector for FD detection from intraoral images. FD-SOS has two novel components: \textbf{conditional contrastive denoising (CCDN)} and \textbf{teeth-specific matching assignment (TMA)}. These modules enable FD-SOS to effectively leverage external dental semantics. Experimental results showed that our method outperformed existing detection methods and surpassed dental professionals by 35\% recall under the same level of precision. Code is available at:~\url{https://github.com/xmed-lab/FD-SOS}.

\end{abstract}
\keywords{Vision Language Model  \and Open-Set Object Detection}
\vspace{-0.2cm}

\input{Section_1_Introduction.tex}

\input{Section_2_Methods.tex}

\input{Section_3_Experiments.tex}

\input{Section_4_Ablation_Conc.tex}

\bibliographystyle{splncs04}
\bibliography{main}

\include{Supp}

\end{document}

%% file: Section_1_Introduction.tex
\section{Introduction}

Bone fenestration and dehiscence (FD) are abnormal conditions that affect the supporting structures of teeth, potentially leading to compromised oral health and tooth loss if left untreated. Achieving accurate detection of FD is crucial for developing effective treatment plans and delivering optimal patient care in dentistry. Currently, dentists primarily rely on analyzing intraoral images of the teeth and gums to make a preliminary diagnosis of FD. However, this diagnostic process often lacks accuracy and suffers from subjectivity. While cone beam computed tomography (CBCT) is considered the gold standard for diagnosing FD in clinical practice, its frequent use in orthodontic treatment, especially in pediatric patients, is associated with radiation-related cancer risks~\cite{Yeh2018EstimatedRR}. Moreover, CBCT scans are costly and less accessible, particularly in low-income countries.
Therefore, there is significant value in developing an FD detection approach that can achieve higher accuracy than dentists while eliminating the need for CBCT scans, using only intraoral images.

\begin{figure}[t]
    \centering    
\includegraphics[width=\textwidth]{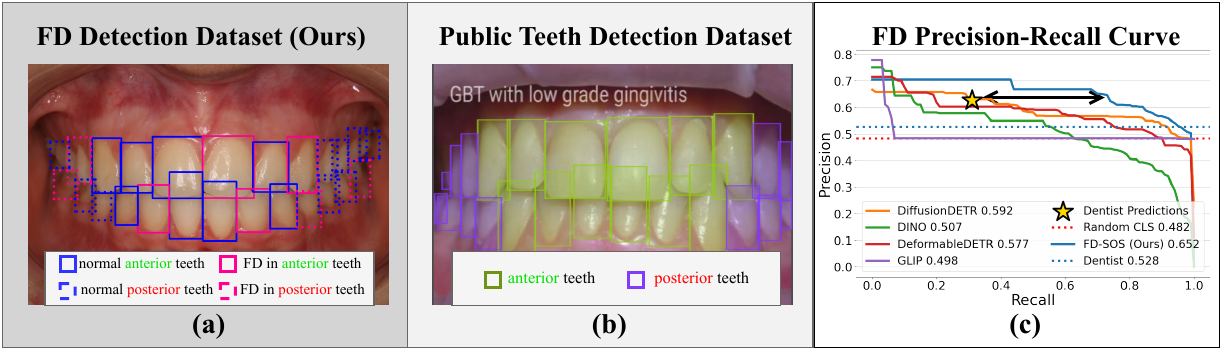}
    \caption{
(a) Our collected dataset for FD detection focuses on anterior teeth without posterior teeth annotations, as indicated by dashed bounding boxes. (b) Publicly available datasets for teeth detection, incorporating two classes. (c) FD detection Precision-Recall Curve shows that our method outperforms other existing detection methods and even professional dentists.  RandomCLS is an always-positive predictor on the true bounding boxes to establish the random average precision (AP).
} 
\label{fig:intro}
\vspace{-0.7cm}
\end{figure}

While there have been deep learning approaches for detecting FD from CBCT scans~\cite{LIU2023105082_FD_CBCT}, no prior work has focused on FD detection from intraoral images. To address this gap, we have collected an in-house dataset called FDTooth, which consists of 150 intraoral images with annotated bounding boxes of anterior teeth provided by dentists, as well as corresponding FD detection results obtained through CBCT.  
As shown in~\cref{fig:intro}(a), the dataset only includes annotations for FD and normal cases in anterior teeth. There are no specific annotations for posterior teeth provided, given FD is not visible in frontal view images. This leads to missing annotations for dashed bounding boxes for posterior teeth. 

FD detection from intraoral images can be considered an object detection task that involves two subtasks: accurately localizing the teeth (``Anterior'' or ``Posterior'') and diagnosing them as either ``No FD'' or ``FD''. One possible solution is to employ object detectors such as DETR~\cite{DETR}, YOLO~\cite{Ge2021YOLOXEY}, FCOS~\cite{tian2021fcos}, or Faster-RCNN~\cite{ren2015faster_rcnn} or their variants like DDETR~\cite{zhu2021deformable}, DINO~\cite{zhang2023dino} and DiffusionDETR~\cite{Chen_2023_ICCV_diffuse_detr}, with multiple heads and multi-label optimization objectives such as binary cross-entropy, Focal~\cite{Lin2017FocalLF}, TMLL~\cite{Kobayashi_2023_CVPR_TMLL}. Yet, directly applying these methods to our dataset for FD detection yields unsatisfactory results; see ``w/o pretraining'' in~\cref{tbl:sota}. This can be primarily attributed to the limited size of our dataset, which poses challenges in terms of enlarging it, as our dataset relies on CBCT as the ground truth for each tooth. Therefore, leveraging publicly available teeth detection datasets, which provide detection labels specifically for anterior and posterior teeth (as shown in~\cref{fig:intro} (b)), for pretraining the network and subsequently fine-tuning it on our dataset, is a potential solution. 
Another potential solution is to develop a multi-task object detection framework by using both the public dataset and our dataset to perform ``anterior/posterior'' teeth detection and ``FD/No FD'' detection, respectively. 
However, despite these approaches, we still observe limited results; see ``Object Detectors $\star$'' in~\cref{tbl:sota}.

\textbf{Vision-Language Models (VLMs)}, trained on vast multimodal datasets, have emerged as Open-Set Object Detectors (OSOD), setting new SOTAs in detection across natural image benchmarks~\cite{zhang2022glipv2,li2021grounded,li2022blip,Li2023BLIP2BL,Liu2023GroundingDM}. For FD detection, fine-tuning OSOD exhibits enhanced performance than standard object detection approaches. Nevertheless, OSOD's effectiveness is constrained, primarily due to catastrophic forgetting and overfitting on limited datasets.

To this end, we present FD-SOS, which stands for~\textbf{FD} \textbf{S}creening through \textbf{O}pen-\textbf{S}et object detectors in intraoral images.
Specifically, we introduce a \textbf{mutli-task VLM} framework that utilizes publicly available dental datasets to enhance FD detection, acting as regularization to mitigate overfitting and improve generalization of VLMs. Given that the public dataset contains a larger number of anterior teeth samples without explicit labels indicating their normal or FD status, we propose a \textbf{novel conditional contrastive denoising (CCDN)} to effectively utilize the shared dental semantic information. Moreover, we introduce \textbf{teeth-specific matching assignment (TMA)} to enhance the model's confidence in case of missing annotations. Through extensive experiments, our FD-SOS 
outperform existing detection methods and \textbf{surpass dental professionals by 35\%} recall under the same level of precision as depicted in~\cref{fig:intro}(c).

%% file: Section_2_Methods.tex
\section{Methodology} 
\begin{table}[t]
\centering
\caption{Comparisons of our FDTOOTH  and Public Teeth Detection dataset from~\cite{Dwyer2024Roboflow}.}
\label{tab:dataset_comparison}
\resizebox{0.6\columnwidth}{!}{

\begin{tabular}{@{}lcc@{}}
\toprule
\textbf{Feature} & \textbf{FDTOOTH (Ours)} & \textbf{Public Teeth Detection} \\
\midrule
Image Num. & 150 & 5,000 \\
Resolution & 5,760 $\times$ 3,840 & 416 $\times$ 416 \\
BBox Annot.
 & Normal, FD; 1,800 & Anterior, Posterior; 20,000 \\
Disease Label & $\checkmark$ & $\times$ \\
Ground Truth & CBCT scans & None \\
Train Split & 90 imgs (454 healthy, 626 FD) & 70\% (3,500 imgs) \\
Val Split & 20 imgs (95 healthy, 145 FD) & 10\% (500 imgs) \\
Test Split & 40 imgs (248 healthy, 232 FD) & 20\% (1,000 imgs) \\

\bottomrule
\end{tabular}}
\end{table}

\subsection{Datasets}

Since there is a lack of publicly available datasets specifically designed for FD detection from intraoral images, we collected a new dataset called \textbf{FDTOOTH}.  Table~\ref{tab:dataset_comparison} shows a comparison between our dataset and the existing public dataset. Notably, annotations for posterior teeth in our dataset are not available since FD in posterior teeth cannot be detected from a frontal intra-oral view. The public dataset only provides bounding box annotations for all teeth, including both anterior and posterior, without any FD-specific information.

\begin{figure}[t]
    \centering
    \includegraphics[width=\textwidth]{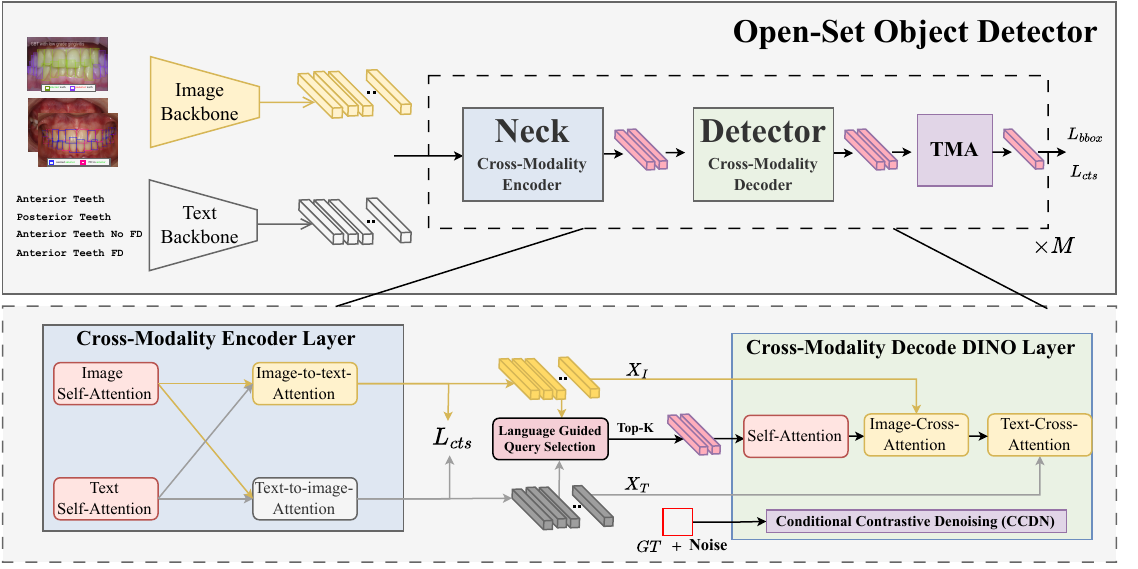}
    \caption{Framework for FD-SOS. }
    \label{fig:grounded_detection}
\end{figure}
\subsection{Overall Framework}

As depicted~\cref{fig:grounded_detection}, we start by extracting text features from each dataset as ``Anterior Teeth'' and ``Posterior Teeth'' for teeth detection and ``Anterior Teeth No FD'' and ``Anterior Teeth FD'' for FD detection. Leveraging Grounding DINO as our VLM baseline~\cite{Liu2023GroundingDM}, we fuse image-text features at various stages, employing a multimodal image-text contrastive loss to enhance the alignment between the visual and textual modalities. Then, these aligned image-text features are used for object localization and classification tasks. A novel Conditional Contrastive Denoising (CCDN) is proposed during the decoding process for the detection head, DINO~\cite{zhang2023dino}, exploiting the common semantics of dental structures available in both datasets (e.g ``Anterior Teeth''). Finally, we leverage a positional prior to introduce a Teeth-specific Matching Assignment (TMA) to mask out positive detections that are missing in the ground truth.

\begin{figure}[t]
    \centering
\includegraphics[width=\textwidth]{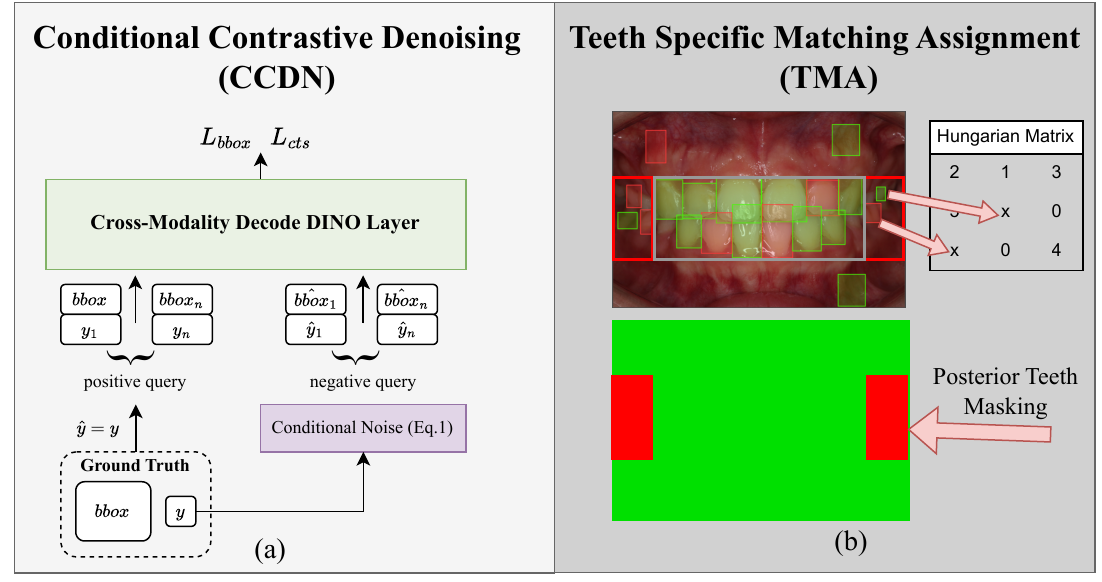}
    \caption{(a) Conditional Contrastive Denoising (CCDN) improves the detection decoder by utilizing attribute-based contrastive denoising. (b) Teeth-Specific Matching Assignment (TMA) leverages positional priors to mask posterior teeth and focus on the differential diagnosis of FD in frontal intra-oral images.} 
\label{fig:TMA}
\end{figure}

\subsection{Conditional Contrastive Denoising}

The ad-hoc approach of encoding the multi-task information in text,  although a promising baseline, falls short by not considering the interconnected label space of dental attributes, thus limiting its effectiveness. For instance, G-DINO~\cite{Liu2023GroundingDM}, an OSOD building on DINO~\cite{zhang2023dino}, employs contrastive denoising to improve training stability. However, its approach to selecting positive and negative anchors for denoising is indiscriminate, overlooking potential spurious correlations caused by overlapping labels across tasks. Specifically, categorizing ``FD'' in undiagnosed ``Anterior teeth'' as a negative query fails to recognize that these teeth may be affected by ``FD'' in a cross-task context. To overcome these challenges, we propose \textbf{a novel conditional contrastive denoising (CCDN)} approach.

Let \( Y \) denote a random variable representing the query label for the denoising task within DINO head~\cite{zhang2023dino}. The label flipping is determined by a probability \( p \) conditioned on the dental attributes of the original label \( y \), namely its diagnosis \( D \) (where \( D = 0 \) indicates ``No FD'' and \( D = 1 \) indicates ``FD''), and position \( L \) (where \( L = 0 \) indicates ``Posterior'' and \( L = 1 \) indicates ``Anterior''). The conditional probability \( P(Y = y' \mid y) \), which defines the likelihood of transitioning from ground truth \( y \) to query label \( y' \) for the denoising task, is given by:

\begin{equation}
P(Y = y' \mid y) =
\begin{cases} 
1 - p & \text{if } y' = y, \\
p & \text{if } y' \neq y \text{ and } L_{y'} \neq L_y, \\
\frac{p}{2} & \text{if } y' \neq y \text{ and } L_{y'} = L_y = 1 \text{ and } D_{y'} \neq D_y, \\
0 & \text{otherwise.}
\end{cases}
\end{equation}

The first condition acts as the \textbf{positive query} for reconstruction depicted in~\cref{fig:TMA} (a). The second condition indicates a position shift (e.g., from ``Posterior'' to ``Anterior'' or vice versa) and is considered a \textbf{negative query}. When there is a diagnosis available within the ``Anterior teeth'', the third condition specifies a change by altering the diagnosis, with a probability of $p/2$ to be considered as a \textbf{negative query}. The fourth condition denotes cases where transitions do not generate noisy queries due to the uncertainty of adding a diagnosis to an undiagnosed tooth. The bounding box ($BBOX$) noise adheres to the baseline head, DINO~\cite{zhang2023dino}. The localization loss, $L_{BBOX}$, follow DETR-like models~\cite{Chen_2023_ICCV_diffuse_detr,zhang2023dino,zhu2021deformable} with L1 and Generalized Intersection Over Union (GIOU) losses. The classification loss leverages contrastive text-image predictions with focal loss~\cite{Lin2017FocalLF}. The contrastive classification loss $L_{cts}$ is defined as:

\begin{equation}
L_{cts} = Focal( X_I \cdot X_T^T, y_{GT})
\end{equation}

Here \(X_I\) and \(X_T\) correspond to the image queries and text features respectively. $ X_I \cdot X_T^T$ is the dot product output logits, and $y_{GT}$ is ground truth label.

\subsection{Teeth-Specific Hungarian Matching Assignment}

Object detection models, including OSOD, utilize a sampling approach for the detection head where positive and negative boxes are sampled for a better learning procedure. These models can be characterized by multi-stage~\cite{9344609_cascade_faster,ren2015faster_rcnn} or single-stage object detectors~\cite{tian2021fcos}, with the latter being more robust to missing annotations~\cite{9506790_sparse_2}. However, sampling negative queries for training in all object detection models can lead to problems when dealing with imperfect and missing annotations~\cite{sissi,Suri_2023_ICCV_sparse_1}. Since the frontal FD detection task is missing annotations within the posterior teeth, it is essential to automatically exclude them during training. To this end, we propose a Teeth-specific Matching Assignment (TMA).

As depicted in~\cref{fig:TMA} (b). we leverage a \textbf{positional prior} about the anterior and posterior teeth, disregarding the posterior teeth during the loss calculation if its information is not available. Utilizing traditional image processing techniques, we identify predicted bounding boxes as posterior teeth if they fall outside the vertical bounds yet lie within the horizontal confines of the anterior teeth region.  

Given an image \(I\) with a set of bounding boxes \(\{(x_1, y_1, x_2, y_2)_b\}\), where \((x_1, y_1)_b\) denotes the top-left corner and \((x_2, y_2)_b\) denotes the bottom-right corner of each bounding box \(b\), we aim to identify the extremities, $E$, of the anterior teeth region. This can be accomplished by defining the following limits:

\begin{equation}
    E_{\text{left}} = \min_{b} (x_1), \quad
E_{\text{right}} = \max_{b} (x_2), \quad
E_{\text{top}} = \min_{b} (y_1), \quad
E_{\text{bottom}} = \max_{b} (y_2)
\end{equation}

We then generate a mask for the posterior teeth region as colored in red in~\cref{fig:TMA} (b) as:

\begin{equation}
mask= (\left( x^{p}_1 < E_{\text{left}} \right) \lor \left( x^{p}_2 > E_{\text{right}} \right)) \land \left( \left( y^{p}_1 > E_{\text{top}} \right) \land \left( y^{p}_2 < E_{\text{bottom}} \right) \right)
\end{equation}

Following, we assign ``$x$'' as a no-care value for predicted bounding boxes within the red mask in the Hungarian matching for object detection~\cite{ren2015faster_rcnn,DETR}, implicitly masking the unannotated true positives during gradient calculation. This positional prior assignment is simple and effectively handles missing posterior teeth annotations, increasing model confidence without the need for complex architecture or pseudo-label mining, focusing on the differential diagnosis of FD in frontal intra-oral images.

%% file: Section_3_Experiments.tex
\section{Experiments}

\noindent{\textbf{Implementation Details.}} For all baselines, we utilize SWIN-T transformer as the image backbone~\cite{liu2021Swin}, which is initialized with ImageNet weights. The models are trained using AdamW optimizer with a batch size of 16. The learning rate is searched within the range of [2.5e-5,5e-4] with mmdetection framework~\cite{mmdetection}. For open-set object detectors, we adopt the pre-trained BERT-base as the textual backbone, which was frozen during the fine-tuning process~\cite{devlin2018bert}.

\noindent{\textbf{Baselines.}} Traditional object detectors include DeformableDETR~\cite{zhu2021deformable}, DINO~\cite{zhang2023dino}, and DiffusionDet~\cite{Chen_2023_ICCV_diffuse_detr} as SOTA object detection models. \textbf{Multi-Label Baselines} include Hierarchical-Diff-DETR~\cite{Hierchiecal_dentex} that utilize DiffusionDet~\cite{Chen_2023_ICCV_diffuse_detr}. Additionally, we utilize all SOTA detectors with the cross-task data employing multi-head objectives. \textbf{Open Set Object detectors baselines (OSOD)}  include GLIP~\cite{li2021grounded} and Grounding DINO~\cite{Liu2023GroundingDM}, SOTA OSOD.

\noindent \textbf{Evaluation Metrics.} Metrics include Pascal VOC ($AP_{50}$) and COCO evaluation ($AP_{75}$, $AP$ averaged over IoU thresholds from 0.5 to 0.95 with a step of 0.05). We report separate results for the positive FD class, resulting in a total of 6 metrics; For all methods, we exclude posterior teeth model output when calculating metrics on the held-out testing set of \textbf{FDTOOTH}. see appendix  for teeth detection (``Anterior'', ``Posterior'') on the cross-task dataset.

\input{Section_3_sota_table}

\subsection{Performance on FD Detection}
\cref{tbl:sota} shows FD detection results on our FDTOOTH test set. A straightforward approach for our task involves applying object detection models without multi-task training. Notably, SOTA object detectors without multi-task training and warmup perform poorly, as illustrated by Diffusion-DETR~\cite{Chen_2023_ICCV_diffuse_detr}, which achieves an $AP_{50}$ of 8.85\%, compared to warming up the model, which significantly improves performance to 66.37\%. 
Incorporating cross-task training in the fine-tuning phase, similar to multi-label objectives as seen in Hierarchical-Diff-DETR~\cite{Hierchiecal_dentex}, leads to reduced performance. This decline is probably attributed to spurious correlations and the absence of annotations, with the model being limited by the initial warmup stage's optimum. Consequently, this results in a decrease in Diffusion-DETR's~\cite{Chen_2023_ICCV_diffuse_detr} $AP_{50}$ from 66.37\% to 64.40\%. An alternative approach for missing annotations is employing SOTA sparse-detection solutions from computer vision. Specifically, SparseDet~\cite{Suri_2023_ICCV_sparse_1} shows enhanced performance achieving $AP_{50}$ of 65.13\%, improvement from Hierarchical-Diff-DETR's 64.40\%. 

\noindent\textbf{Vision-Language Models (VLMs)}, trained on extensive multimodal data and fine-tuned for the FD task without multi-task training, show varied performance. GLIP~\cite{li2021grounded} achieves an $AP_{50}$ of 55.85\%, and GDINO~\cite{Liu2023GroundingDM} achieves 65.89\%, reflecting the detrimental effects of spurious correlations present in multi-task frameworks. GLIP struggles with missing annotations due to its use of full-level text attention. In contrast, GDINO's sub-sentence attention mechanism offers an implicit solution to missing annotations, outperforming GLIP and establishing it as a stronger baseline with an improvement of 10\% on $AP_{50}$. Nonetheless, introducing a multi-task framework does not improve GDINO's performance, likely due to spurious correlations and the shared dental semantic and label space. 
\textbf{FD-SOS} address these challenges establishing new SOTA benchmarks across all metrics. Specifically, \textbf{FD-SOS} surpass the SOTA object detector Diffusion-DETR~\cite{Chen_2023_ICCV_diffuse_detr} by  \textbf{6.9\%} and outperform SparseDet~\cite{Suri_2023_ICCV_sparse_1} by \textbf{11\%} on $AP$, a strict detection metric. Importantly, FD-SOS improves the best-performing \textbf{VLM},  GDINO~\cite{Liu2023GroundingDM}, by \textbf{3.89\%} on the $AP$ metric and \textbf{6.09\%} on the $AP_{FD}$ metric.

\begin{table}[t]
\centering
\caption{Ablation of proposed components.}
\resizebox{0.8\columnwidth}{!}{
\begin{tabular}{c|c|c|ccc|ccc}
\toprule
\multirow{2}{*}{} & \multirow{2}{*}{TMA} & \multirow{2}{*}{CCDN} & \multicolumn{6}{c}{Metrics} \\
\cline{4-9} & & & $AP_{75FD}$ & $AP_{FD}$ & $AP_{50FD}$ & $AP_{75}$ & $AP$ & $AP_{50}$ \\
\hline

GDINO (Baseline) & $\times$ & $\times$ & 55.55 & 54.75 & 59.99 & 62.6 & 62.08 & 65.81 \\
w/ TMA & \checkmark & $\times$ & 59.68 & 58.29 & 63.09 & 64.56 & 63.63 & 67.12 \\
w/ CCDN & $\times$ & \checkmark & 58.59 & 57.08 & 61.83 & 64.34 & 63.41 & 66.87 \\
\textbf{FD-SOS (ours)} & \checkmark & \checkmark & \textbf{62.45} & \textbf{60.84} & \textbf{66.01} & \textbf{67.07} & \textbf{65.97} & \textbf{69.67} \\

\bottomrule
\end{tabular}}
\label{tbl:ablation_gdino}
\end{table}

\noindent \textbf{Ablation Study.} ~\cref{tbl:ablation_gdino} shows the ablation results of our FD-SOS. Using our proposed conditional contrastive denoising (CCDN) increases the performance across all metrics, specifically by 2.3\% on $AP_{FD}$. The increase is solely attributed to the removal of spurious correlation during the denoising process, enhancing the performance of the baseline and extending it to multi-task detection. Using only the teeth-specific matching increases the performance by 3.5\% on $AP_{FD}$, indicating that OSOD is still not fully robust to missing annotations, limited by the usage of the detection head, DINO~\cite{zhang2023dino}. Utilizing TMA and CCDN together substantially improves the performance for FD by \textbf{6.9\%} in $AP_{75FD}$, \textbf{6.09\%} in $AP_{FD}$, and \textbf{6.02\%} in $AP_{50FD}$ compared to the baseline, GDINO~\cite{Liu2023GroundingDM}.

\begin{figure}[h]
    \centering
\includegraphics[width=\textwidth]{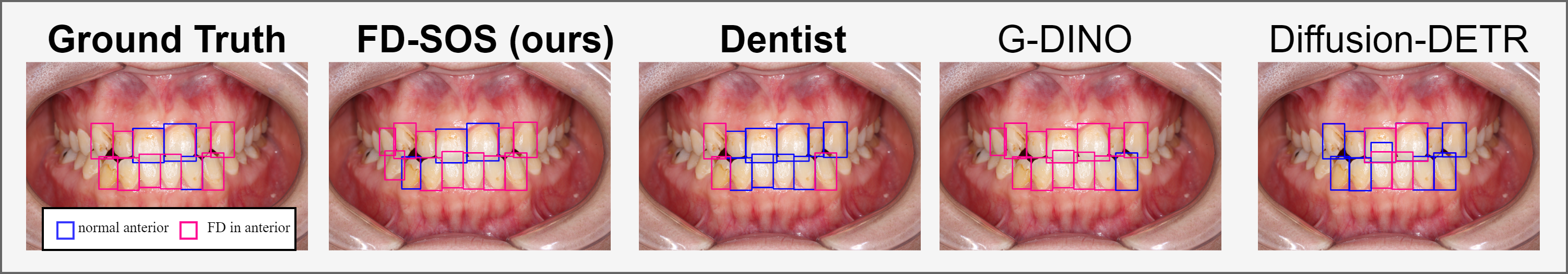}
    \caption{Qualitative results for the best-performing methods on FD detection.}
    \label{fig:qualitaive}
\end{figure}

%% file: Section_3_sota_table.tex
\begin{table}[t]
\centering
\caption{Results of FD detection on our dataset. All object detectors commence with initialization from ImageNet pre-trained weights. $\star$ requires pre-training on the publicly dental dataset~\cite{Dwyer2024Roboflow}. $\dagger$ refers to fine-tuning existing VLM pre-trained models. FD-SOS shares the same complexity as the baseline G-DINO (64.02M), higher than DINO (48.04M), and significantly lower than GLIP (122.8M). 
}
\label{tbl:sota}
\resizebox{1.0\columnwidth}{!}{
\begin{tabular}{l|c|ccccc|c}
\toprule
   Methods  & multi-task & $AP_{75FD}$ & $AP_{FD}$ & $AP_{50FD}$ & $AP_{75}$ & $AP$ & $\boldsymbol{AP_{50}}$ \\
  \hline
  \multicolumn{8}{c}{Object Detectors~$\star$} \\
  \hline
  Diffusion-DETR  w/o pretraining~\cite{Chen_2023_ICCV_diffuse_detr}  & $\times$ & 0.04 & 1.31 & 7.58 & 0.04 & 1.7 & 8.85 \\
  Diffusion-DETR~\cite{Chen_2023_ICCV_diffuse_detr}  & $\times$ & 55.52 & 51.42 & 61.28 & 62.58 & 59.06 & 66.37 \\
  DDETR~\cite{zhu2021deformable}  & $\times$ & 56.92 & 50.41 & 60.51 & 62.68 & 57.44 & 65.48 \\
  DINO~\cite{zhang2023dino}  & $\times$ & 54.03 & 49.68 & 57.94 & 55.13 & 51.65 & 57.65 \\
  \hline
  Hierarchical-Diff-DetR w/o pretraining~\cite{Hierchiecal_dentex}  & $\checkmark$ & 0.0 & 0.0 & 0.0 & 0.0 & 0.0 & 0.0 \\
  Hierarchical-Diff-DetR~\cite{Hierchiecal_dentex}  & $\checkmark$ & 55.81 & 50.82 & 59.21 & 61.45 & 57.63 & 64.4 \\
  DDETR~\cite{zhu2021deformable} & $\checkmark$ & 53.28 & 47.57 & 58.01 & 60.2 & 54.89 & 63.05 \\
  DINO~\cite{zhang2023dino}  & $\checkmark$ & 48.12 & 43.85 & 50.69 & 53.8 & 50.15 & 55.53 \\
  SparseDet~\cite{Suri_2023_ICCV_sparse_1} & $\checkmark$ & 60.5 & 52.94 & 62.95 & 62.96 & 54.87 & 65.13 \\
  \hline
  \multicolumn{8}{c}{Open-set Object Detectors~$\dagger$} \\
  \hline
  GLIP~\cite{li2021grounded}  & $\times$ & 40.57 & 32.0 & 46.34 & 51.3 & 40.47 & 55.85 \\
  GDINO~\cite{Liu2023GroundingDM}  & $\times$ & 58.32 & 56.59 & 61.07 & 63.69 & 62.59 & 65.89 \\
  \hline
  GLIP~\cite{li2021grounded}  & $\checkmark$ & 41.78 & 33.68 & 47.09 & 51.97 & 42.73 & 56.7 \\
  GDINO~\cite{Liu2023GroundingDM} (our baseline) & $\checkmark$ & 55.55 & 54.75 & 59.99 & 62.6 & 62.08 & 65.81 \\
  \textbf{FD-SOS (ours)}  & $\checkmark$ & \textbf{62.45} & \textbf{60.84} & \textbf{66.01} & \textbf{67.07} & \textbf{65.97} & \textbf{69.67} \\
\bottomrule
\end{tabular}}
\end{table}

%% file: Section_4_Ablation_Conc.tex
\section{Conclusion}

This paper presents a clinically significant application - FD detection from intraoral images. We introduce FD-SOS, a novel OSOD-based framework, which incorporates two innovative components: CCDN and TMA. The results show that our method significantly outperforms existing detection methods and even surpasses the performance of dental professionals, highlighting its immense clinical value in practical settings. Furthermore, this method can be extended to other dental diseases once corresponding datasets become available.

\begin{credits}
\subsubsection{\ackname} 
This work was supported in part by grants from the National Natural Science Foundation of China under Grant No. 62306254, grants from the Research Grants Council of the Hong Kong Special Administrative Region, China (Project Reference Number: T45-401/22-N), and Project of Hetao Shenzhen-Hong Kong Science and Technology Innovation Cooperation Zone (HZQB-KCZYB-2020083).
Marawan Elbatel is supported by the Hong Kong PhD Fellowship Scheme (HKPFS) from the Hong Kong Research Grants Council (RGC), and by the Belt and Road Initiative from the HKSAR Government.
\subsubsection{\discintname}
The authors have no competing interests to declare that are relevant to the content of this article.
\end{credits}

%% file: Supp.tex
\appendix

\section*{\centering{Appendix for ``FD-SOS''}}

\begin{figure}[h]
    \centering
\includegraphics[width=0.8\textwidth]{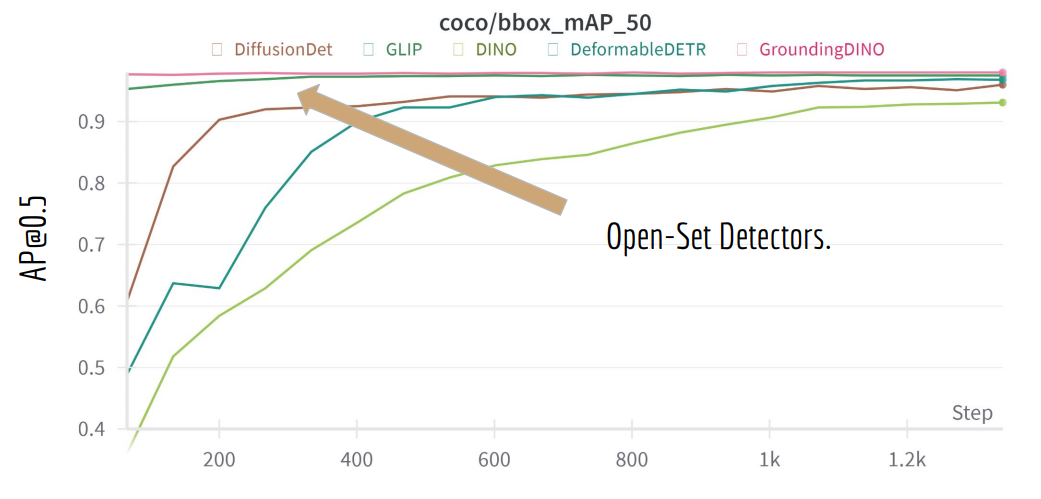}
\caption{Teeth detection is relatively an easier task. Results on the validation set for the cross-task teeth detection, encompassing both ``Anterior'' and ``Posterior'' categories, demonstrate that open-set detectors do not require a warmup phase, as they are already familiar with the task of teeth detection from the outset, unlike other detection models such as  DeformableDETR, DINO, or DiffusionDETR models. The achieved performance for teeth detection, with an \(AP_{50} > 90\%\) indicates that it is effectively a solved problem in the field of computer vision.}
\label{fig:lbl}
\end{figure}